\documentclass[conference]{IEEEtran}
\IEEEoverridecommandlockouts
\usepackage{cite}
\usepackage{amsmath,amssymb,amsfonts}
\usepackage{algorithmic}
\usepackage{graphicx}
\usepackage{textcomp}
\usepackage{xcolor}
\def\BibTeX{{\rm B\kern-.05em{\sc i\kern-.025em b}\kern-.08em
    T\kern-.1667em\lower.7ex\hbox{E}\kern-.125emX}}
\begin{document}

\title{Real Time Control and Performance Analysis \\ of
	A Smart Multi-Machine System through 
	Hardware-in-the-Loop Simulation
}

\author{
	\IEEEauthorblockN{Muhammad Sarwar$^1$, Anas Ramzan$^2$, Muhammad Usman Naseer$^3$, Bilal Asad$^4$}
	\IEEEauthorblockA{\textit{$^{1-4}$Dept. of Electrical Engineering} \\
		\textit{$^1$Pakistan Institute of Engineering \& Applied Sciences}, 
		Islamabad, Pakistan \\
		\textit{$^{2-4}$The Islamia University of Bahwalpur}, 
		Bahawalpur, Pakistan \\		
		$^1$msarwar@pieas.edu.pk, $^2$m.anas13@ymail.com, $^3$usman.naseer@iub.edu.pk, $^4$bilal.asad@iub.edu.pk}
}

\maketitle

\begin{abstract}
This paper presents an implementation of a smart power system using inter-device communication and supervisory control through Simulink integrated with the physical system. This control includes features of real-time control, load management, fault detection and self-healing. The prototype is designed in hardware which takes in required parameter via sensors to PC via RS-232 power meter interface and actuator interfaced with Arduino microcontroller board which is integrated with Simulink. Simulink simulates the model and predict control signals depending upon input parameters and it sends control signal back to Arduino which then control the hardware through hardware in the loop (HIL) simulation. In the case of a severe fault, these control techniques detect system variation and compensate for this variation or remove the faulty part by using relays, which represents the self-healing nature of the developed model.
\end{abstract}

\begin{IEEEkeywords}
— Smart grid (SG), HIL, Simulink, Power system, Micro grid, Smart control
\end{IEEEkeywords}

\section{Introduction}
Electricity is the best form of energy that can  be utilized to cope with the modern needs of a progressive civilization. Electrical energy generated at power stations is transferred to consumers through a complex transmission and distribution network known as power grid. Currently, electric power is produced in a centralized manner, in which it is transmitted from generating stations to load centers in a uni-directional hierarchical flow \cite{farhangi2010path}. This unidirectional, uncontrolled flow of electric power poses a number of challenges to grid operators, thus jeopardizing the security, quality and reliability of the electric power being supplied at the consumer’s end \cite{kundur1994power}.

The future grid that features bidirectional transfer of power from central generation centers to consumers by deploying information and communication technologies and also from Distributed Generators (DGs) to other consumers and controls all the processes in a pervasive and smart manner can be termed as Smart Grid (SG) \cite{brown2001analyzing}. Furnished with state-of-the-art communication and smart control infrastructure, the future Smart Grid will successfully be able to cope with increasing demand of the bulk loads and will prevail the power industry market with economic and environmental advantages \cite{ipakchi2009grid}. A thorough review of the challenges and technologies for the future smart grid is given in \cite{sarwar2016review} for interested readers.

To materialize dream of such an advanced grid, conventional power system needs to be reinvented at all levels. Its aging infrastructure and less efficient networks need to be replaced by flexible and digitally smart equipment. Such a network level overhaul requires an industrious effort of technical research and huge investments in modern state-of-the-art technologies to evolve them into the grid of future. An effort has been made in this survey to uncover modern research being carried out in this field, its benefits, investors and companies interested in revamping the traditional grid to serve humanity in a much better way.

The conventional grid is prone to structural weaknesses and raises hazardous environmental concerns. Theses weaknesses and concerns have severe impact on reliability and quality of the power provided by conventional grid. Hence, there is a great need of advance and sustainable grid that uses modern information technologies for secure, reliable dispatch of electricity both from bulk generating stations and from Distributed Energy Resources (DERs) in a multi-directional and flexible manner with exchange of information in real time \cite{hassan2010survey}.

\begin{figure}
	\centerline{\includegraphics[width=0.5\textwidth]{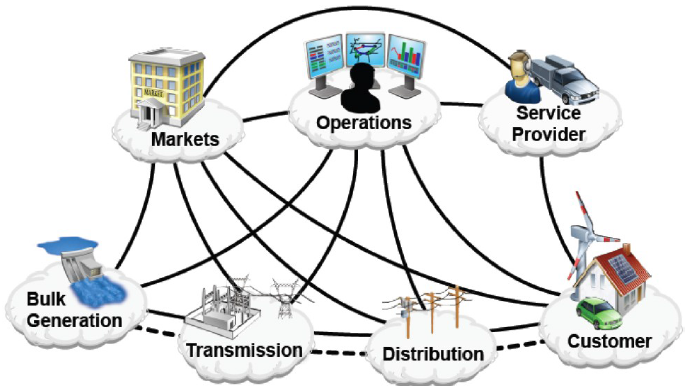}}
	\caption{A Conceptual Model for Future Smart Grid}
	\label{fig1}
\end{figure}

Efficiency of a power system decreases due to the losses in it. Similarly, the instability of multi-machine depends upon the transients produced during its operation. To overcome these losses a micro grid model should be assembled in such a way that all the controlled parameters should apply on it \cite{amin2005toward}. The results of these parameters should have the ability to apply on macro grids and with the help of these parameters control on the losses is possible. In our proposed model, we use two motor generator set as a generating station and 3 buses for load.

A LabVIEW based test-bed system has been designed by authors of \cite{mahmood2019design} where they designed and test an automatic synchronizing and protection relay for the distribution generator (DG). A smart micro-grid system is designed by the authors of \cite{abubakar2018real} in which different protection and automation techniques and algorithms have been implemented to enhance the security and reliability of the designed micro-grid.

Objective of this paper is to simulate and develop the model in such a way that all the controlled parameters i.e. power flow analysis, damping of transients, improvements in power factor, smart load sharing, and system stability are applicable on it. And also capable of smart power generation and distribution including automatic load shedding, forecasting and multi-agent system for inter-device communication.

A working model of SG is presented in this paper that comprises of advanced technologies, networks and subsystems. The model describes SG as a system of systems through a systematic approach from modern technologies and innovative engineering practices towards the grid of future. We also implement future grid concept to make grid automated and self-healing under the widespread control of ultra-smart management and control systems providing a number of benefits to both utilities and consumers. Being aware of huge landscape of the SG, future grid will be comprised of smart infrastructure, smart information, communication, management and protection \cite{hashmi2011survey}.

In Section II, the prototype implementation is described and it’s working according to different load condition for the system. In Section III, discussion and implementation of prototype in SIMULINK are explained while results and analysis are discussed in the Section IV. In Section V, a conclusion about more stabilized and automated future power system by using more advance technologies is highlighted.

\section{Prototype Implementation}
Prototype is implemented in two ways:

\subsection{Software Implementation}
For the control and monitoring of the power system, input and output interfaces and control algorithms are developed in Simulink which takes inputs from system in real time and control the system parameters through its output interface. Arduino is also integrated with MATLAB for faster and precise numerical computations of the load forecasting and reliability assessment of the system.

\subsection{Hardware Implementation}
Smart Generation and distribution system is developed and interfaced with software model. Hardware model consists of generators with control equipment and smart distribution system. Distribution system consists of resistive and inductive loads, solid state control circuits which are used to distribute and control power in a smart manner.

Following steps should be followed, first of all make a motor-generator set. For this purpose, connect the synchronous machine with tachometer. Also connect the available D.C motor of suitable rating with same tachometer. Here we use D.C motor as a prime-mover. Purpose of connection of D.C motor as well as synchronous machine with tachometer is to determine the speed, torque and power of these machines at any instant of time Tachometer has two parts, one is called Sensor Unit while second is called Display Unit. Sensor Unit is connected with motor and generator shaft for sake of sensing and determining the change in speed and torque of the motor and generator. These determined values will display on Display Unit with the help of serial cable connected with Sensor Unit.

\begin{figure*}
	\centerline{\includegraphics[width=\textwidth]{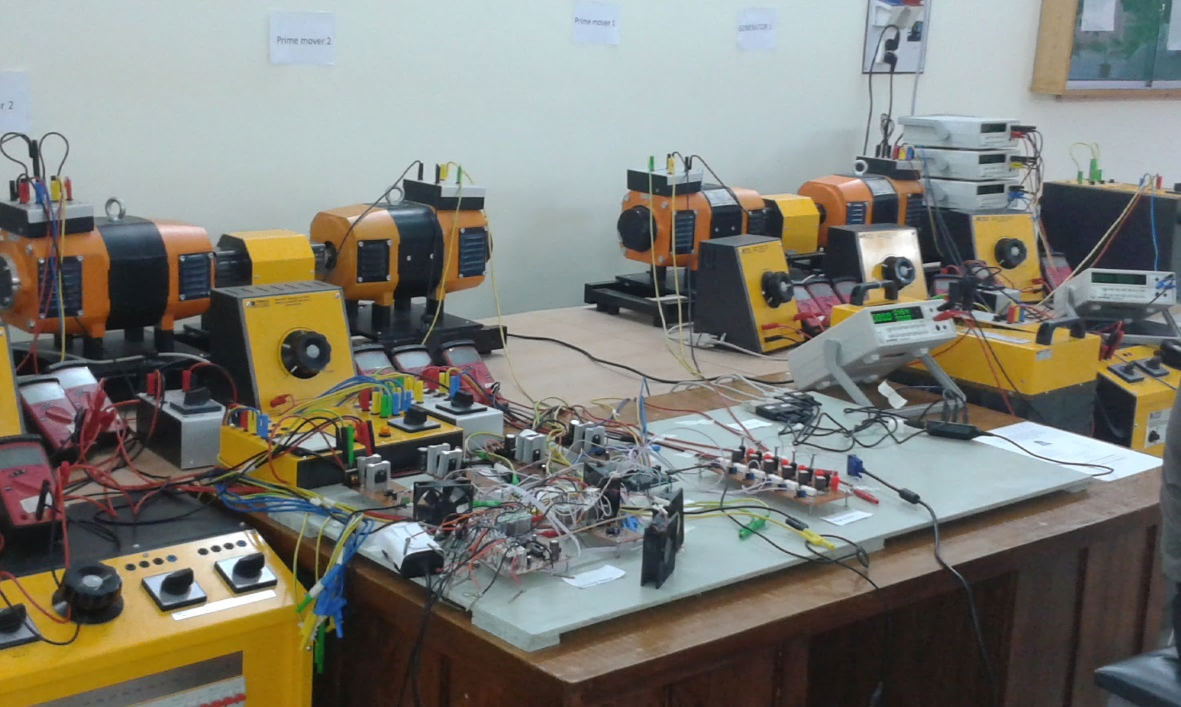}}
	\caption{Hardware implemented model}
	\label{fig2}
\end{figure*}

Output of generator is connected with 3-phase A.C power meter to determine the output voltages of generator. And Block diagram is given in fig 3. The aim is to automate the same system and control it with computer or by any other type of software based device. Use Buck Converter in-place of rheostat. To control all these circuits such as Buck Converter, Current Sensor and Voltage Sensor, use software based Arduino which is connected to operating system such as computer via its software driver.

Arduino is a programmable device which has number of I.Cs. The limiting values of parameters i.e. maximum and minimum values of current, voltage of motor-generator set is pre-programmed in Arduino. So that the maximum and minimum range of parameters of motor-generator system should not exceed. If there is any change in values occur, Arduino gets signal of that change and it corrects that undesirable change and stable the output value to its pre-programmed value. Basic purpose of buck Converter is to regulate the D.C source supply so that the input to field could be controlled and varied. Buck Converter get pulse with modulated signal through Arduino.

\section{Block Diagram and Component Specification}
\begin{figure}
	\centerline{\includegraphics[width=0.5\textwidth]{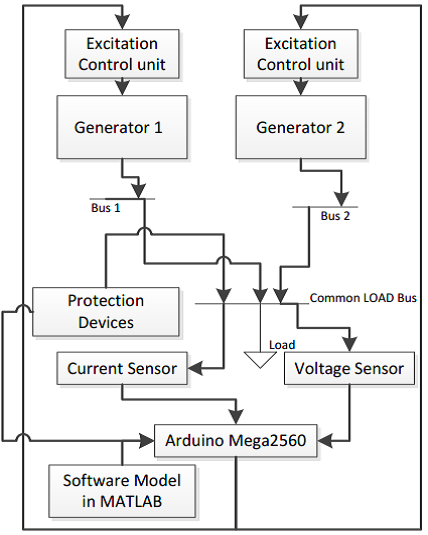}}
	\caption{Block diagram for Control of HIL Model}
	\label{fig3}
\end{figure}

To monitor and its stable operation, our prototype follows following control structure in hardware and software implementation. Both generator feed power to the load through Bus1 and Bus 2 to common load bus and generator change its speed and other parameter according to load on the common load bus. Control structure sense variation in load current and voltage through current and voltage sensors respectively and different protection devices i.e. relay and breaker to common load bus. These variation feed directly to Arduino and in the same way, model developed in MATLAB also working on the same specifications. Shown in Fig. \ref{fig3}.
Arduino work as a communication channel between hardware and software model and integrate them in such a way that they respond each other in any variation.

Arduino Compare the MATLAB values and hardware model and generate signal to make system stable. Like, in case of failure of any generating station load shift on healthy generating station, Arduino sense the change and increase/decrease the excitation of generator to meet the load demand and remove faulty generator through circuit breaker. And similarly load variation cause different abnormal behaviors, current sensor measure and communicate with Arduino to check the current limit. If it is in permissible range then it will generate the signal and change the excitation of generator by changing the duty cycle of buck converter to meet load variation. If current variation is not in permissible range then it will generate signal to remove the extra load to make it stable.

\begin{figure}
	\centerline{\includegraphics[width=0.5\textwidth]{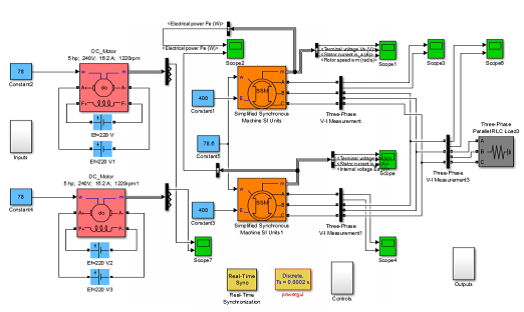}}
	\caption{ Simulink Model of the physical system}
	\label{fig4}
\end{figure}

In this way, both hardware and software interlink with each other and communicate to make the system self-healing and whole system performance can be consider as a hardware in the loop (HIL) simulation.

\section{Simulink Model}
In software implementation observer can observe the effects of change in values of parameters while output of motor-generator set remains same. Voltage Sensor and Current Sensor controls and limits the value of applied voltage and current, to avoid the excessive flow of current in the field winding of motor. Voltage Sensor and Current Sensor measures the input line current and limit this value to a definite ratio or some pre-determined ratio and gives signal to Arduino device. Arduino compare these receiving values with its pre-programmed value and perform the necessary change in value so as to keep the stable output value. Real-time output value is displayed on the screen of software based device such as computer etc. Graphical representation of change in values can be displayed on screen for graphical analysis \cite{cirio2007extensive}.

We basically integrate hardware model with software model which takes the input from hardware model and compare it with Simulink design model which is working under the ideal condition and error signal is generate by the Arduino then feed into prime mover to change the system performance. When system undergoes from the severe fault then this part can automatically remove or try to move it towards its stable state. Rating of different component is shown in Table \ref{table1}.

\begin{table}
	\caption{Rating of Different Component in HIL Model}
	\begin{center}
		\begin{tabular}{|c|c|}
			\hline
			\textbf{Component Name } & \textbf{Ratings } \\ \hline 
			Arduino Due & 3.3V \\ \hline 
			Generating station & 1.2kW, 1400 RPM \\ \hline 
			Power pack & 3.5A, 220V,230V, 10A(3-phase) \\ \hline 
			Torque power meter & 3000 RPM, +/- 5.5kW, +/-17.5Nm \\ \hline 
			Compound motor & 1kW, 220V Excitation(0.55 A) \\ \hline 
			Load switch & 3-pole, 16A,250V DC/440V AC \\ \hline 
			Voltage Sensor & 3.3 V \\
			\hline
		\end{tabular}
		\label{table1}
	\end{center}
\end{table}

\begin{figure}
	\centerline{\includegraphics[width=0.5\textwidth]{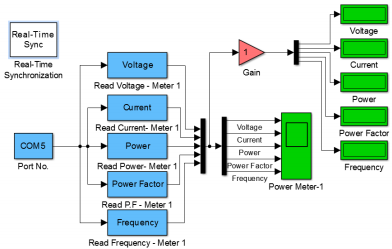}}
	\caption{Power meter RS-232 communication protocol through Simulink}
	\label{fig5}
\end{figure}

Prototype is designed in hardware which takes in required parameter via sensors to PC via RS-232 power meter interface and actuator interfaced with Arduino microcontroller board which is integrated with MATLAB Simulink shown in Fig. \ref{fig5}.

We observe our system under different condition like, timing variation, frequency variation, and many more. We can estimate timing behavior because under this observation we are able to observe the system stability and also system response under various condition by using MATLAB Simulink. The main purpose of hardware in the loop is that we run the prototype simultaneously in software and hardware at a time and do the analysis by observing the results showing in graph and time response is about 1ms. We can improve it by introducing more sensitive devices.

\begin{figure}
	\centerline{\includegraphics[width=0.5\textwidth, height=10cm]{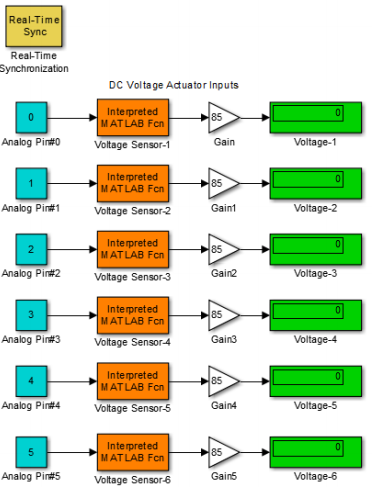}}
	\caption{System input voltage sensor model}
	\label{fig6}
\end{figure}

System voltage variation is actuated by MATLAB Simulink model shown in Fig. \ref{fig6}.

\section{Analysis and Results}
The system when completely synchronized with the Simulink model, all the inputs from physical systems are passed on the virtual system using actuators and sensors. AC power meter is used to transfer Electrical parameters of the generators and loads to the virtual model. DC voltage and current sensors pass on the DC voltage and current values to the system running in the Simulink. Torque and tachometer unit displays the values of shaft torque, RPM and power produced by each prime mover thus keeping a check on each prime mover’s real power input to the generator.

\begin{figure}
	\centerline{\includegraphics[width=0.5\textwidth,height=10cm]{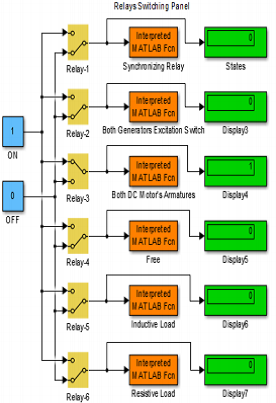}}
	\caption{Relay Control Panel}
	\label{fig7}
\end{figure}

The system controls the hardware parameters as well. Switching relays switch on and off the load and shed the load when high fault currents are passed through the system. DC voltage regulators (buck converter) control the excitation of each alternator and field voltage of each prime mover (DC motor) which is connected as separately excited motor. The system in up and running condition control the speed of each prime mover, voltage of each generator, automatically synchronize the two alternators when commanded from Simulink model, reads the electrical parameters of all the generators, motors and loads attached.

\begin{figure}
	\centerline{\includegraphics[width=0.5\textwidth,height=7cm]{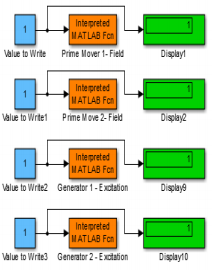}}
	\caption{Buck converter duty cycle control}
	\label{fig8}
\end{figure}

From system performance, we can observe that terminal voltage, stator current and rotor speed is constant at 78.7. But when load on a generator 1 increase then rotor speed also fluctuates and this effect also shifts towards generator 2. But after comparing it with our MATLAB model, it will try to move generator speed towards a constant rpm, internal voltage, stator current and all other parameter start to stabilize. And load current and voltage is also constant. Fig9-10 represents system performance and behavior under stable condition.

\begin{figure}
	\centerline{\includegraphics[width=0.5\textwidth]{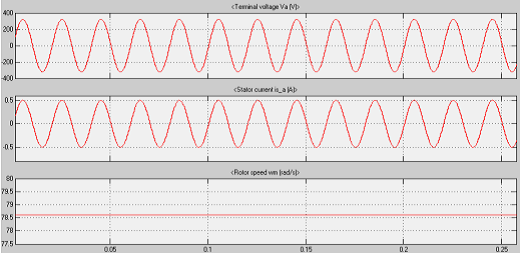}}
	\caption{Outputs of Generator-1}
	\label{fig9}
\end{figure}

\begin{figure}
	\centerline{\includegraphics[width=0.5\textwidth]{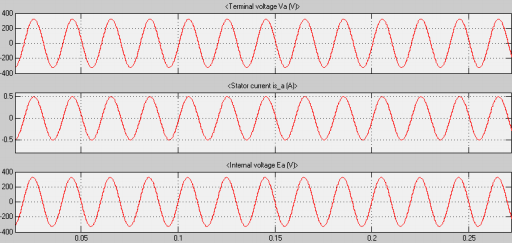}}
	\caption{Terminal Outputs of Generator-2}
	\label{fig10}
\end{figure}

\begin{figure}
	\centerline{\includegraphics[width=0.5\textwidth]{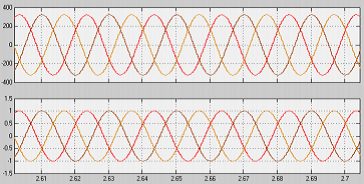}}
	\caption{Voltage \& Current waveforms at Load Terminals}
	\label{fig11}
\end{figure}

\section{Conclusion}
Future grid is an ineludible technological innovation that would help us to build an environmental friendly and sustainable future for energy demands. This paper reviews major pilot projects initiated by various utilities, organizations and institutions. And tried to present a self-automated system, which undergoes different system variations and then stabilizes according to the system needs. The risks introduced due to deployment of advanced technologies were also assessed and solutions proposed by various researchers were presented. The system can be extended to a more advanced and reliable system that would incorporate a 3 machine 9 bus system with IEEE standard bus data. For the same prototype, the Simulink model can be extended into a GUI control panel which displays all the values of input parameters, provided controls for all variable parameters and provides options for system analysis and control.

\bibliographystyle{IEEEtran}

\bibliography{bsfypRef}

\begin{thebibliography}{10}
\providecommand{\url}[1]{#1}
\csname url@samestyle\endcsname
\providecommand{\newblock}{\relax}
\providecommand{\bibinfo}[2]{#2}
\providecommand{\BIBentrySTDinterwordspacing}{\spaceskip=0pt\relax}
\providecommand{\BIBentryALTinterwordstretchfactor}{4}
\providecommand{\BIBentryALTinterwordspacing}{\spaceskip=\fontdimen2\font plus
\BIBentryALTinterwordstretchfactor\fontdimen3\font minus
  \fontdimen4\font\relax}
\providecommand{\BIBforeignlanguage}[2]{{%
\expandafter\ifx\csname l@#1\endcsname\relax
\typeout{** WARNING: IEEEtran.bst: No hyphenation pattern has been}%
\typeout{** loaded for the language `#1'. Using the pattern for}%
\typeout{** the default language instead.}%
\else
\language=\csname l@#1\endcsname
\fi
#2}}
\providecommand{\BIBdecl}{\relax}
\BIBdecl

\bibitem{farhangi2010path}
H.~Farhangi, ``The path of the smart grid,'' \emph{IEEE power and energy
  magazine}, vol.~8, no.~1, 2010.

\bibitem{kundur1994power}
P.~Kundur, N.~J. Balu, and M.~G. Lauby, \emph{Power system stability and
  control}.\hskip 1em plus 0.5em minus 0.4em\relax McGraw-hill New York, 1994,
  vol.~7.

\bibitem{brown2001analyzing}
R.~E. Brown and L.~A. Freeman, ``Analyzing the reliability impact of
  distributed generation,'' in \emph{Power Engineering Society Summer Meeting,
  2001}, vol.~2.\hskip 1em plus 0.5em minus 0.4em\relax IEEE, 2001, pp.
  1013--1018.

\bibitem{ipakchi2009grid}
A.~Ipakchi and F.~Albuyeh, ``Grid of the future,'' \emph{IEEE power and energy
  magazine}, vol.~7, no.~2, pp. 52--62, 2009.

\bibitem{sarwar2016review}
M.~Sarwar and B.~Asad, ``A review on future power systems; technologies and
  research for smart grids,'' in \emph{2016 International Conference on
  Emerging Technologies (ICET)}.\hskip 1em plus 0.5em minus 0.4em\relax IEEE,
  October 2016, pp. 1--6.

\bibitem{hassan2010survey}
R.~Hassan and G.~Radman, ``Survey on smart grid,'' in \emph{IEEE SoutheastCon
  2010 (SoutheastCon), Proceedings of the}.\hskip 1em plus 0.5em minus
  0.4em\relax IEEE, 2010, pp. 210--213.

\bibitem{amin2005toward}
S.~M. Amin and B.~F. Wollenberg, ``Toward a smart grid: power delivery for the
  21st century,'' \emph{IEEE power and energy magazine}, vol.~3, no.~5, pp.
  34--41, 2005.

\bibitem{mahmood2019design}
M.~Mahmood, M.~Azam, K.-u.-N. Fatima, M.~Sarwar, M.~Abubakar, and B.~Hussain,
  ``Design and implementation of an automatic synchronizing and protection
  relay through power-hardware-in-the-loop (phil) simulation,'' \emph{arXiv
  preprint arXiv:1907.00339}, 2019.

\bibitem{abubakar2018real}
M.~{Abubakar}, M.~{Sarwar}, T.~H. {Khan}, M.~{Sarmad}, and B.~{Hussain},
  ``Real-time implementation of directional over-current protection
  coordination in a microgrid,'' in \emph{2018 International Symposium on
  Recent Advances in Electrical Engineering (RAEE)}, Oct 2018, pp. 1--6.

\bibitem{hashmi2011survey}
M.~Hashmi, S.~H{\"a}nninen, and K.~M{\"a}ki, ``Survey of smart grid concepts,
  architectures, and technological demonstrations worldwide,'' in
  \emph{Innovative Smart Grid Technologies (ISGT Latin America), 2011 IEEE PES
  Conference on}.\hskip 1em plus 0.5em minus 0.4em\relax IEEE, 2011, pp. 1--7.

\bibitem{cirio2007extensive}
D.~Cirio, E.~Gaglioti, S.~Massucco, A.~Pitto, and F.~Silvestro, ``An extensive
  dynamic security assessment analysis on a large realistic electric power
  system,'' in \emph{Power Tech, 2007 IEEE Lausanne}.\hskip 1em plus 0.5em
  minus 0.4em\relax IEEE, 2007, pp. 285--292.

\end{thebibliography}

\end{document}